\documentclass[11pt]{article}  %Do not change the point size.
\usepackage{menuproc}
\usepackage{cite}
\usepackage{epsfig}
\usepackage{amsmath,amssymb}
\begin{document}
%
%____________________________________________________________
%
%  Title, authors, institutions, and abstract
%----------------------------------------------------------------
%  Syntax:  \titlematter{title}{authors}{institutions}{abstract}
%----------------------------------------------------------------
%     If lines are too long, use linebreaks where convenient.
%     If all authors are from the same institution, omit raised letters.
%
\titlematter{Study of the $\eta$-proton interaction via the reaction $pp\rightarrow pp\eta$}%
{P.~Moskal$^{a,b}$,
 H.-H.~Adam$^c$,
 A.~Budzanowski$^d$,
 R.~Czy{\.{z}}ykiewicz$^a$,
 T.~G{\"o}tz$^b$,
 D.~Grzonka$^b$,
 L.~Jarczyk$^a$,
 A.~Khoukaz$^c$,
 K.~Kilian$^b$,
 C.~Kolf $^b$,
 P.~Kowina$^{b,e}$,
 N.~Lang$^c$,
 T.~Lister$^c$,
 W.~Oelert$^b$,
 C.~Quentmeier$^c$,
 R.~Santo$^c$,
 G.~Schepers$^b$,
 T.~Sefzick$^b$,
 M.~Siemaszko$^e$,
 J.~Smyrski$^a$,
 S.~Steltenkamp$^c$,
 A.~Strza{\l}kowski$^a$,
 P.~Winter$^b$,
 M.~Wolke$^b$,
 P.~W{\"u}stner$^b$,
 W.~Zipper$^e$}%
{$^a$ Institute of Physics, Jagellonian University,  Cracow, Poland \\
 $^b$ IKP $\&$ ZEL, Forschungszentrum J\"{u}lich, Germany \\
 $^c$ Institut f{\"u}r Kernphysik,
                    Westf\"{a}lische Wilhelms--Universit\"{a}t, M\"unster, Germany\\
 $^d$ Institute of Nuclear Physics, Cracow, Poland \\
 $^e$ Institute of Physics, University of Silesia,  Katowice, Poland}%
{ A  measurement of the $pp\rightarrow pp\eta$ reaction
  at the excess energy of Q~=~15.5~$\pm$~0.4~MeV  has been carried out at 
  the internal beam facility COSY-11 with an integrated  luminosity of  811~nb$^{-1}$.
  The number of $\sim$24000 identified events  permits a precise determination 
  of total (2.32~$\pm$~0.05~$\pm$~0.35~$\mu b$) 
  and differential cross sections.
  Preliminary investigations show that the  
  angular distribution of the $\eta$ meson in the center-of-mass system
  is isotropic. 
  A qualitative analysis of the Dalitz-plot distribution is presented.
}
%
%
%____________________________________________________________
%  Start article here:

%%%%%%%%%%%%%%%%%%%%%%%%%%%%%%%%%%%%%%%%%%%%%%%%%%%%%%%%%%%%%%%%%%%%%%%%%%%%%%%%%%
\section{Introduction}
Investigations of the $\eta$ meson production  via the
$pp \rightarrow pp\eta$ reaction address the question
of the strength of the proton-$\eta$ interaction at low relative momenta
of the interacting particles. In the frame of the optical potential
model this interaction
can be expressed in terms of phase shifts, which in turn are
described by the scattering length $a_{\eta N}$ and the effective
range of the potential.
Usually,  the $a_{\eta N}$  is defined as a complex quantity
with the imaginary part accounting for the $\eta N \rightarrow \pi N$
and $\eta N \rightarrow \pi \pi  N$ processes.
The real part of it
is a direct measure
of the formation -- or non-formation -- of an $\eta$-nuclear
quasi-bound state~\cite{svarc}.
At present 
it is still not known whether
the  attractive interaction between $\eta$ meson and nucleons 
is strong enough to form an $\eta$-mesic nucleus
or a quasi-bound $\eta$NN state.
The values of Re(a$_{\eta N}$) range between 0.25~fm and 1.05~fm depending 
on the analysis method and the studied reaction~\cite{wycechgreen}.
According to reference\cite{rakityansky}, 
within the present inaccuracy of Re(a$_{\eta N}$) 
the existence of quasi-bound  $\eta$-mesic light nuclei could be possible.
The shape of the energy dependence of the $pd \to ^3$He\,$\eta$ cross section
implies that either the real or imaginary part of the $\eta \, ^3\mbox{He}$ scattering
length has to be very large~\cite{wilkin93}, which may be associated with 
a bound $\eta \, ^3\mbox{He}$ system.
Similarly encouraging are results of reference\cite{shevchenko},
where it is argued that
a three-body $\eta$NN resonant state, which may be formed
close to the $\eta$d threshold, may evolve into
a quasi-bound state for Re($a_{\eta N}$)~$\ge$~0.733~fm.
Also the close to threshold enhancement
of the total cross section of the
$pp \rightarrow pp\eta$ reaction\cite{etadata}
was interpreted as being
either a Borromean (quasi-bound) -- or a resonance  $\eta$pp state~\cite{wycech2},
provided that  Re($a_{\eta N}$)~$\ge$~0.7~fm.
Contrary, recent calculations performed within
a three-body formalism indicate~\cite{fix} that a formation
of a three-body $\eta$NN resonance state is rather not
possible, independently of the $\eta$N scattering parameters.
Moreover, the authors of reference~\cite{garcilazo1} exclude
the possibility of the existence of an $\eta$NN quasi-bound state.
 However, results of both calculations~\cite{fix,garcilazo1},
although performed within a three-body formalism, used the assumption
of a separability of the two-body $\eta$N and NN interactions, and
hence the new quality in the three-body  $\eta$NN-interaction
is not excluded and deserves experimental investigations. \\

\section{Experimental results}
 A Close to threshold measurement of the $pp \to pp\eta$ reaction allows
 to study the interaction of the $\eta$-meson with the proton.
 At an excess energy of Q~=~15.5~MeV, at which the reported measurement
 has been performed, 
 the final state particles are in the range of the
 strong interaction much longer than $10^{-23}$~s -- typical 
 life-time of N$^*$ and $\Delta$ baryon resonances.
 Thus their mutual interaction may significantly influence 
 the distributions of their relative momenta.
 
 By means of the COSY-11 detection system\cite{brauksiepe},
 using a stochastically cooled proton beam 
 of the cooler synchrotron COSY\cite{prasuhn} 
 and a hydrogen cluster target\cite{dombrowski},
 we have performed a high
 statistics measurement of the $pp \rightarrow pp\eta$ reaction
 at an excess energy of Q~=~15.5~MeV.
 The experiment was based on the four-momentum registration 
 of both outgoing protons, whereas the $\eta$ meson was identified
 via the missing mass technique.
 Figure~1a presents the missing mass spectrum,
 with the clear signal originating from  $\sim$24000
 events of the $pp\rightarrow pp\eta$ reaction seen
 on a flat distribution due to  multi-pion production.
 By means of the simultaneous measurement of elastically scattered protons
 we were able to monitor  not only the luminosity but 
 also the synchrotron beam geometrical dimensions and its position
 relative to the target\cite{monitorNIM}.
 This, and the correction for the mean beam-momentum-changes
 determined by means of the Schottky-spectrum and the known beam optics,
 allow us to reproduce exactly the observed missing mass
 distribution 
 as it is shown by the dashed
 line in Figure~1a, which is hardly distinguishable from the real data.
 Figure~1b shows that the full range of the $\eta$ meson
 center-of-mass polar scattering angles has been covered by the
 detection system acceptance. This permitted to determine the 
 angular distribution of the created $\eta$ meson
 which, as can be seen in Figure~2a, is completely isotropic
 within the shown statistical errors. The observed distribution 
 is consistent with the previous measurement performed 
 at an excess energy of Q~=~16~MeV at  the CELSIUS facility~\cite{etahigher}.
 However, it 
 improves the former statistics by a factor of 80.
 The determination of the four-momentum vectors for both
 outgoing protons of each registered event gives the complete
 information of the $\eta$pp-system allowing for
 investigations of the $\eta$p and $\eta$pp interactions.
 Figures~2b and 2c show the Dalitz-plots of the identified
 $pp\eta$ system corrected for the detection acceptance
 and the proton-proton interaction.
 The enhancement from the $\eta$-proton
 interaction at small m$_{p\eta}^2$ is evident.
 However, one can also easily
 recognize a difference between Figures~2b and 2c, which originates 
 from various prescriptions of the proton-proton FSI enhancement factors.
 It is well established that for the close-to-threshold meson production
 the energy dependence of the total cross section and the distributions
 of the differential cross section are predominantly determined
 by the nucleon-nucleon final state interaction~\cite{fsiproc}.
 However,  when reducing the proton-proton FSI effect to a multiplicative factor,
 one finds that 
 it depends on the assumed nucleon-nucleon 
 potential and
 on the produced meson mass~\cite{baruFSI}. 
 Figures 2b and 2c present 
 the extreme cases in the estimation of the proton-proton
\vspace{-0.8cm}
\begin{figure}[h!]
\parbox{.33\textwidth}{\centerline{\epsfig{file=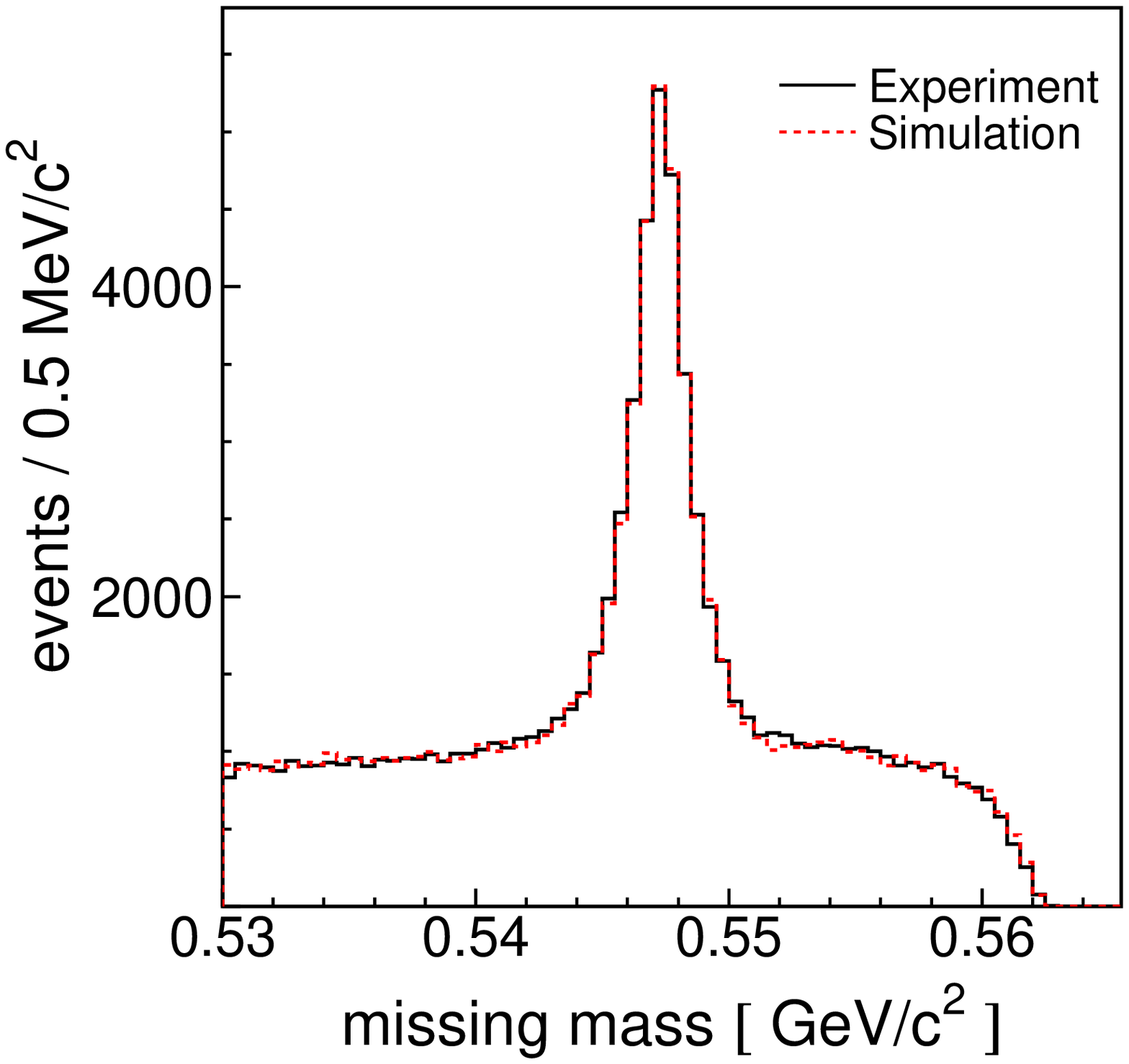,
                      width=.37\textwidth,silent=,clip=}}}%
\parbox{.33\textwidth}{\centerline{\epsfig{file=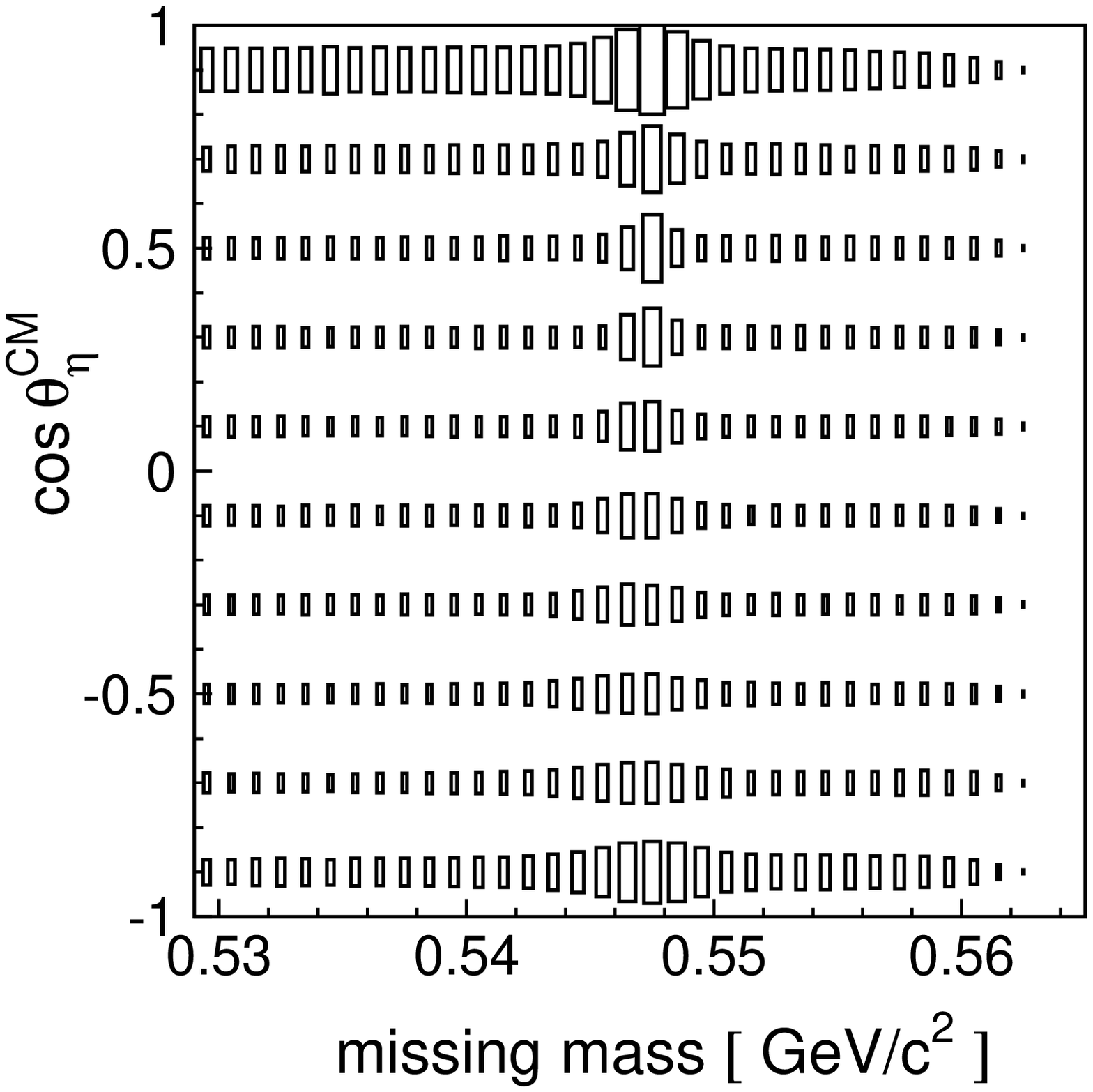,
                      width=.37\textwidth,silent=,clip=}}}%
\parbox{0.33\textwidth}
       {\caption{\label{figure1}
          (a) Missing mass spectrum for the $pp\rightarrow ppX$
           reaction determined 
           %%by means of the COSY-11 detection system
           at a beam momentum of 2.0259~GeV/c.
           The mass resolution amounts to 1~MeV/c$^2$~($\sigma$). \ \
          (b) Distribution of the  center-of-mass polar angle of the produced
           system X as a function of the missing mass.
       }}

\vspace{-0.15cm}
\parbox{.27\textwidth}{\mbox{}}
\parbox{.04\textwidth}{(a)}
\parbox{.28\textwidth}{\mbox{}}
\parbox{.04\textwidth}{(b)}
\end{figure}
\begin{figure}[t!]
\vspace{-1.3cm}
\parbox{.33\textwidth}{\centerline{\epsfig{file=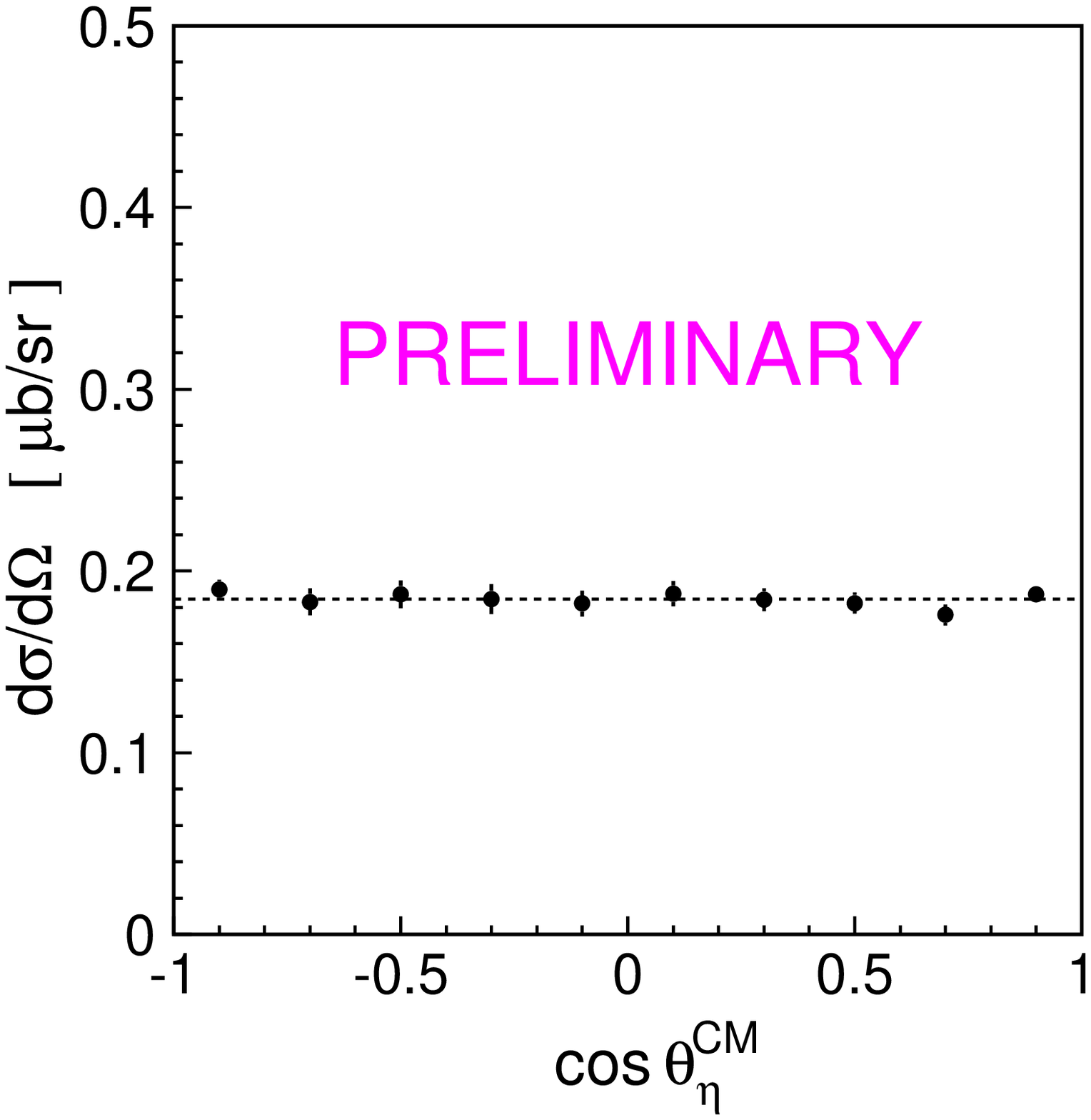,
                      width=.37\textwidth,silent=,clip=}}}%
\parbox{.33\textwidth}{\centerline{\epsfig{file=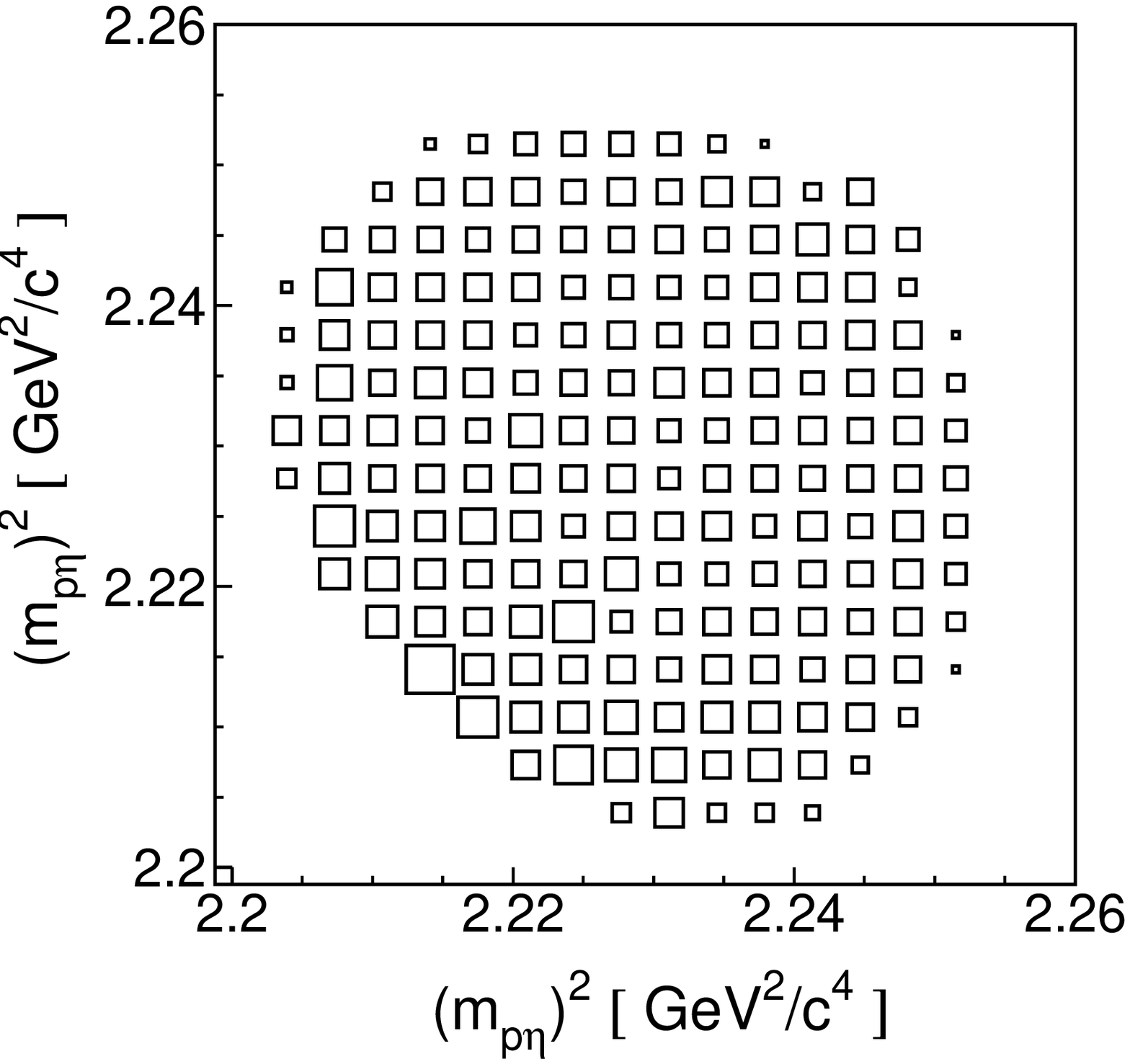,
                      width=.37\textwidth,silent=,clip=}}}%
\parbox{.33\textwidth}{\centerline{\epsfig{file=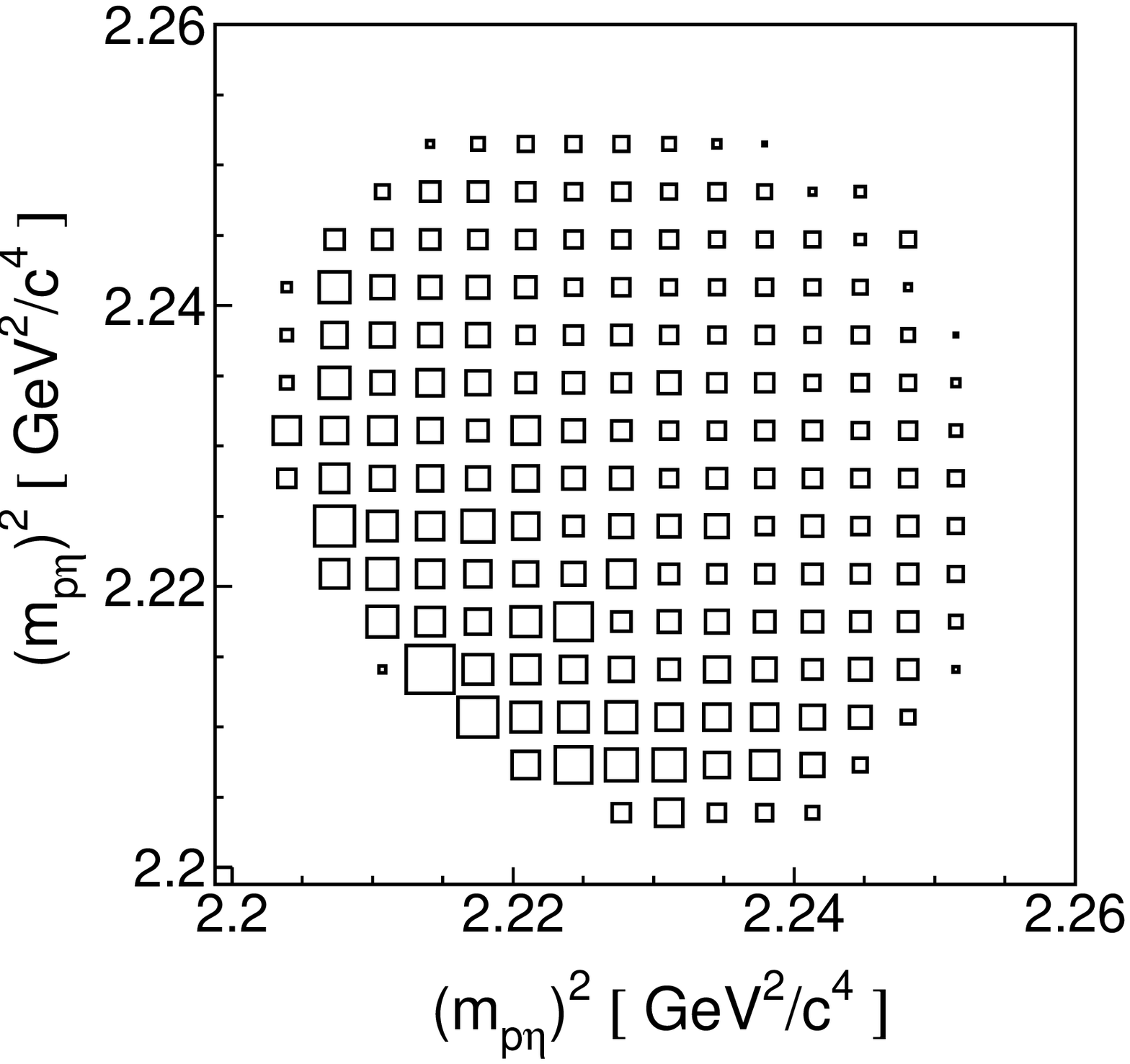,
                      width=.37\textwidth,silent=,clip=}}}%
\vspace{-0.4cm}
\parbox{.28\textwidth}{\mbox{}}
\parbox{.04\textwidth}{(a)}
\parbox{.28\textwidth}{\mbox{}}
\parbox{.04\textwidth}{(b)}
\parbox{.28\textwidth}{\mbox{}}
\parbox{.04\textwidth}{(c)}
\caption{\label{figure2}
         (a) Differential cross section of the $pp \to pp \eta$ reaction 
           as a function of the $\eta$ meson center-of-mass polar angle.
         (b) Dalitz-plot distribution corrected for the
           detection acceptance and the proton-proton FSI.
           For this plot only events with a mass differing
           by no more than 1~MeV/c$^2$ from the
           real $\eta$ meson mass were taken into account.
           The proton-proton FSI enhancement factor was calculated
           as an inverse of the Jost function presented in reference\cite{nisk}.
         (c) The same as (b) but the enhancement factor accounting
           for the proton-proton FSI was calculated as a square of the
           on-shell proton-proton scattering amplitude
           derived according to the modified Cini-Fubini-Stanghellini
           formula including Wong-Noyes Coulomb corrections~\cite{swave,noyes}.
        }
\end{figure}
 FSI effects\cite{swave}. Due to these differences a derivation of the $\eta p$ or
 $\eta pp$ scattering length from the taken data will require a careful estimation
 of the model dependence of corrections for the proton-proton FSI.
\vspace{-0.4cm}
%%%%%%%%%%%%%%%%%%%%%%%%%%%%%%%%%%%%%%%%%%%%%%%%%%%%%%%%%%%%%%%%%%%%%%%%%%%%%%%%%%
\acknowledgments{
  This work was partly supported by the European Community - Access to Research
  Infrastructure action of the Improving Human Potential Programme.
}
%%%%%%%%%%%%%%%%%%%%%%%%%%%%%%%%%%%%%%%%%%%%%%%%%%%%%%%%%%%%%%%%%%%%%%%%%%%%%%%%%%
\vspace{-0.5cm}
%____________________________________________________________
%  Start references here:

\end{document}